\documentclass[fleqn,10pt]{wlscirep}

\usepackage{mathtools}

\title{Thermal noise and optomechanical features in the emission of a membrane-coupled compound cavity laser diode}

\author[1,2,+,*]{Lorenzo Baldacci}
\author[1,+]{Alessandro Pitanti}
\author[1]{Luca Masini}
\author[1]{Andrea Arcangeli}
\author[1]{Francesco Colangelo}
\author[1,3]{Daniel Navarro-Urrios}
\author[4]{Alessandro Tredicucci}
\affil[1]{NEST, CNR Istituto Nanoscienze and Scuola Normale Superiore, Piazza San Silvestro 12, 56127, Pisa (Italy)}
\affil[2]{Institute of Life Sciences, Scuola Superiore Sant'Anna, Piazza Martiri della Libert\`a 33, 56127, Pisa (Italy)}
\affil[3]{Catalan Institute of Nanoscience and Nanotechnology (ICN2), CSIS and the Barcelona Institute of Science and Technology, Campus UAB, Bellaterra, 08193, Barcelona (Spain)}
\affil[4]{ NEST, CNR Istituto Nanoscienze and Dipartimento di Fisica, Universit\`a di Pisa, Largo Pontecorvo 3, 56127, Pisa (Italy)}

\affil[*]{l.baldacci@sssup.it}

\affil[+]{these authors contributed equally to this work}


\begin{abstract}
We demonstrate the use of a compound optical cavity as linear displacement detector, by measuring the thermal motion of a silicon nitride suspended membrane acting as the external mirror of a near-infrared Littrow laser diode. Fluctuations in the laser optical power induced by the membrane vibrations are collected by a photodiode integrated within the laser, and then measured with a spectrum analyzer.
The dynamics of the membrane driven by a piezoelectric actuator is investigated as a function of air pressure and actuator displacement in a homodyne configuration. The high Q-factor ($\sim 3.4\cdot 10^4$ at $8.3 \cdot 10^{-3}$ mbar) of the fundamental mechanical mode at $\sim 73$ kHz guarantees a detection sensitivity high enough for direct measurement of thermal motion at room temperature ($\sim 87$ pm RMS).
The compound cavity system here introduced can be employed as a table-top, cost-effective linear displacement detector for cavity optomechanics.
Furthermore, thanks to the strong optical nonlinearities of the laser compound cavity, these systems open new perspectives in the study of non-Markovian quantum properties at the mesoscale.
\end{abstract}
\begin{document}

\flushbottom
\maketitle
%
%
\thispagestyle{empty}


\section*{\label{sec:intro}Introduction}

The fast progress of cavity optomechanics has produced extraordinary advances in fabrication and characterization of nanomechanical objects\cite{aspelmeyer2014rev}, with impressive results towards the achievement of quantum effects in mesoscopic systems \cite{chan2011laser,teufel2011sideband,brooks2012non,wollman2015,groeblacher2016}. Routinely dealing with sub-picometer displacements, the measure of an optomechanical system often requires complicated and expensive setups composed by optical interferometers and local oscillators to perform homodyne or heterodyne experiments. In this paper we show how a table-top, well known vibrometric technique such as the laser self-mixing\cite{Giuliani2002laser,giuliani2002self,otsuka2002real} can be pushed to measure the thermal motion of a typical optomechanical device, that is a metallic mirror mounted on a trampoline membrane\cite{kleckner2011tramp,norte2015tramp,reinhardt2015tramp}. By shining a laser beam back into its resonant cavity after interacting with the external environment, an interferometric system is created which does not require any complicated arrangement or frequency demodulation. Showing the capability of detecting tens of pm displacements in its linear, stable regime, this technique would allow for detection of thermal noise up to cryogenic temperatures, making it an easy to implement, fast characterization benchmark for optomechanical resonators. Moreover, by reaching the unstable dynamical regime, a rich and interesting physics can be accessed, which falls in the almost unexplored topic of active cavity optomechanics \cite{yang2015laser,czerniuk2014}. Using the chaotic dynamics readily attainable in self-mixed lasers\cite{Kane2005undocking,uchida2012optical} scenarios in which chaotic mechanical objects are created through radiation pressure effects can be easily imagined.

\section*{\label{sec:exp_setup}Experimental setup}
\subsection*{\label{sec:setup} Optical bench and signal analysis}
\begin{figure}
\centering
\includegraphics[width = \linewidth]{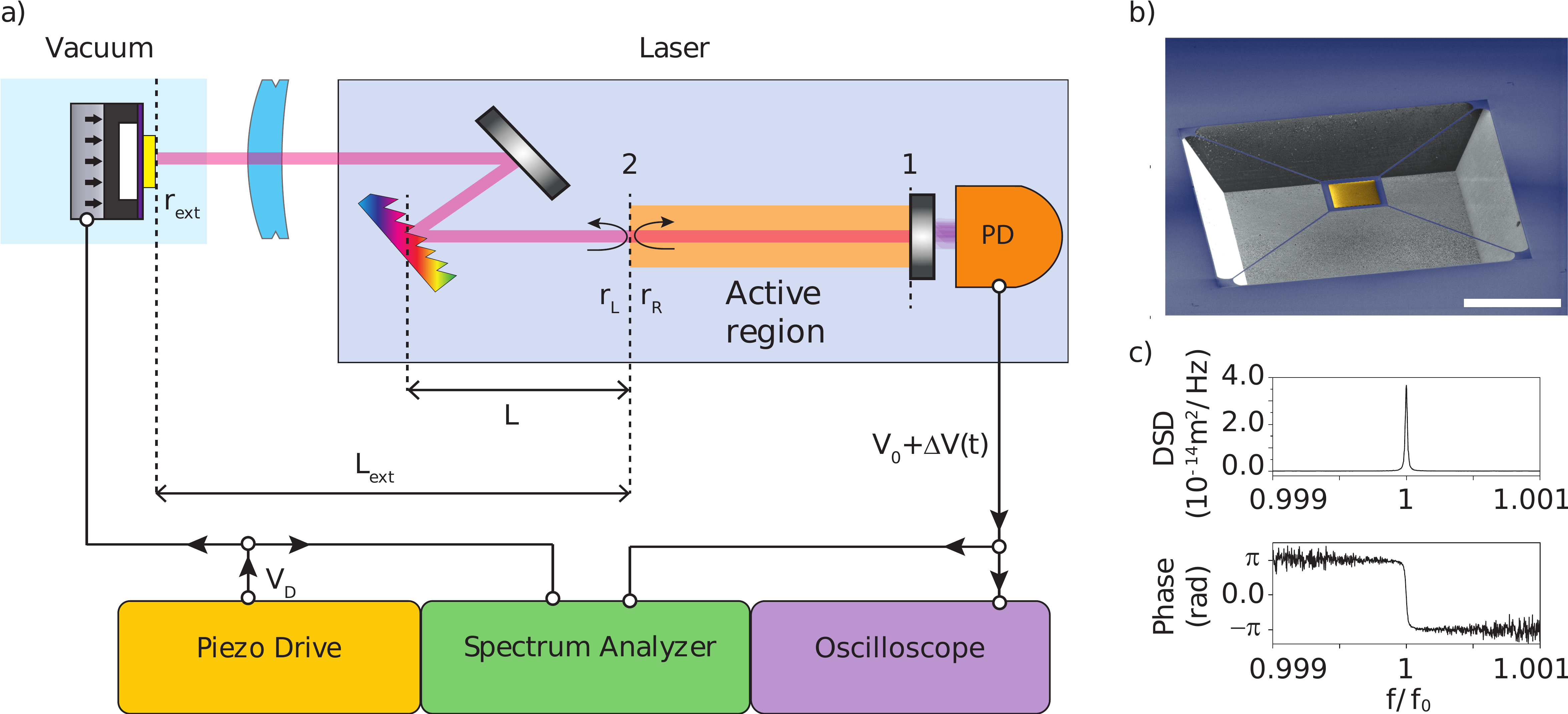}
\caption{\label{fig:setup} a) Sketch of the setup. The laser block is composed by: an active region (light orange bar in the picture), with an anti-reflection coated facet (labeled as 2); an external Littrow grating (multi-colored milled bar) placed at distance $L$ from the facet 2; a coupling mirror (grayscale bar near the Littrow grating); an integrated photodiode (labeled as PD). The emitted radiation is focused by a lens (sky blue curved bar) onto an external mirror (yellow bar) placed at distance $L_{ext}$ from the facet 2, and then reflected back into the laser in order to form a compound cavity. Within a traveling wave approximation, the lasing conditions can be found as function of the effective right and left reflectivities calculated at the facet 2 interface. The radiation emitted by the facet labeled as 1 is collected by the integrated photodiode, and the resulting voltage $V_0+\Delta V(t)$ is recorded by an oscilloscope and a spectrum analyzer.
In the main experiment the external mirror is a gold layer deposited onto a silicon nitride membrane, which is mounted on a piezo actuator in order to drive its displacement by a voltage $V_D$. The resulted device is placed inside a vacuum chamber in order to control the environment pressure. When the membrane is displaced along the optical axis, the output signal is modulated according to eqs. \eqref{eqn:detoma}.
b) SEM image of the Si$_3$N$_4$ membrane. The yellow-colored square at the center of the membrane is the deposited gold layer. The white bar is $200\, \mu \textit{m}$.
c)  The spectrum analyzer reports the power spectral density, and the total amount of displacement can be obtained through proper calibration. An example of measurement is reported for the membrane moved by the piezo actuator. The homodyne approach enables to collect the two motion quadratures of the membrane.}
\end{figure}
The optomechanical laser setup is sketched in Fig. \ref{fig:setup}(a). The optical bench is a compound cavity working as a displacement sensor \cite{Giuliani2002laser,giuliani2002self,otsuka2002real,Kane2005undocking}. The source includes a Littrow external cavity diode laser (ECDL), with lasing wavelength $\lambda_L\simeq 945\, \text{nm}$. The light-emitting, active medium is coupled to a Littrow angle reflection grating, placed at a distance $L = 1$ cm from the left facet of the diode.
The zeroth diffraction order is reflected outside the ECDL and focused by an achromatic doublet (focal length $75 \,\text{mm}$) onto a gold mirror placed on top of a silicon nitride trampoline membrane \cite{kleckner2011tramp,reinhardt2015tramp,norte2015tramp}. The membrane is mounted on a piezo ceramic actuator and placed inside a vacuum chamber, in order to drive its motion and control the environment pressure when needed. The membrane is then aligned to the optical path in such way that light is reflected back into the ECDL to form a compound cavity. As the membrane moves, it changes the oscillation condition of the cavity. Normally this would simply cause a frequency shift in the lasing modes of the cavity: in our case, thanks to the feedback interferometric effect this translates into a modulation of the field amplitude in the laser cavity \cite{Giuliani2002laser,giuliani2002self,otsuka2002real,Lang1980external}.
The resulting laser radiation emitted from the right active region (AR) facet (labeled as 1 in Fig. \ref{fig:setup}(a)) is collected by an integrated photodiode (labeled as PD in Fig. \ref{fig:setup}(a)), which returns an electric potential proportional to the emitted power. For a fixed current bias, this potential, called readout voltage in the following, is composed by a static part $V_0$, and a dynamic term $\Delta V(t)$. $V_0$ is measured with an oscilloscope, while $\Delta V(t)$ is acquired by a spectrum analyzer and read out in the frequency domain. This is the essential quantity to be measured in order to demonstrate the working principle of the setup: by recording the modulation in the laser emitted power, informations about the motion of the membrane are recorded.
A micrograph of the silicon nitride trampoline membrane employed in this work is shown in Fig. \ref{fig:setup}(b). A square-shaped, 5/45 nm Ti/Au layer is deposited on top of the 200 nm nitride, in order to improve its reflectivity. The first resonant mechanical mode oscillates orthogonally to the mirror surface, bending the tethers while keeping the central square parallel to the substrate. The mode frequency is $f_0 =73279.0\pm0.3\,\text{Hz}$ with a quality factor $Q = 34000 \pm 3000 $ (at 0.0083 mbar), limited by both squeeze film effect \cite{blech1983isothermal,bao2007squeeze}, and thermoelastic damping of the four tethers \cite{schmid2011damping}.
When the motion of the membrane is driven by a phase-coherent external displacement source, such as a piezo-ceramic actuator, the two quadratures of motion can be measured through the readout voltage. A typical measurement of amplitude and phase spectrum is shown in Fig. \ref{fig:setup}(c).

\subsection*{\label{sec:littrow}Connection between readout voltage and displacement: theoretical prediction}
In this paragraph the relation between the readout voltage and the external mirror displacement is calculated within the semiclassical framework, in order to predict the sensitivity of the apparatus.
We consider a model system like the one sketched in Fig. \ref{fig:setup}(a). The Littrow ECDL is characterized essentially by the optical and electronic properties of the AR and the optical properties of the blazed angle grating \cite{Ye2004tunable}. In the traveling wave approximation the laser oscillation condition can be expressed with the following set of coupled equations \cite{tromborg1997mode,detoma2005complex}:
\begin{subequations}
\label{eqn:detoma}
\begin{gather}
\label{subeq:detomafield} r_R(\omega)r_L(\omega, N, P) = 1, \\
\label{subeq:detomacarrier}\frac{I-I_0}{q} =  V_c \frac{N-N_{th}}{\tau_s} + \frac{c_0}{n(\omega, N,  P)} g (\omega, N, P) P.
\end{gather}
\end{subequations}
The coefficients $r_R$ and $r_L$ are related to the amplitude of the right and left traveling wave at the AR facet interface, labeled as 2 in Fig. \ref{fig:setup}(a).
The effective index $n$ and the modal gain $g$ are in principle spectral functions of the laser carrier density $N$ and the photon number $P$; the other parameters employed are defined in Table \ref{tab:param}.
Equations \eqref{eqn:detoma} can be solved numerically to find $N$, $P$ and the lasing frequency $f_L$ of the cavity modes, while the stability analysis for small signals \cite{detoma2005complex} will determine which modes are currently lasing.
The ECDL is coupled to an external mirror (i.e. the Si$_3$N$_4$ membrane) with reflectivity $r_{ext}$, which directly affects the $r_R$ coefficient: this becomes a function of the delay time $\tau_{ext}=2L_{ext}/c_0$:
\begin{equation}
r_{R}(\omega, \tau_{ext}) = \frac{r_2+r_G(\omega)e^{-i\omega \tau}+r_2 r_G(\omega) r_{ext}e^{-i\omega \tau_{ext}}+r_{ext}e^{-i\omega (\tau+\tau_{ext})}}{1+r_2 r_G(\omega) e^{-i\omega \tau}+r_G(\omega) r_{ext} e^{-i\omega \tau_{ext}}+r_2 r_{ext} e^{-i\omega (\tau + \tau_{ext})}}.
\label{eqn:rR}
\end{equation}
When the external mirror is moving, $r_{R}$ is modified, changing the solution of eq. \eqref{eqn:detoma} as well. 
The cavity mode radiation envelope evolves with two different timescales: $T_L$, defined by the laser relaxation damping, and $T_M$, defined by the oscillation frequency of the membrane as $T_M = 1/f_0 (\sim 13\, \mu s)$.
We numerically solved eq. \eqref{eqn:detoma} with the parameters relative to our ECDL and found $T_L \sim 10$ ns for the lasing modes, roughly three orders of magnitude smaller than $T_M$. In this case the e.m. radiation envelope can be assumed to evolve through states which verify eq. \eqref{eqn:detoma} instantaneously\cite{mezzapesa2015nanoscale}. This allows us to solve eq. \eqref{eqn:detoma} for different static displacements of the membrane around its equilibrium position. For each position we can define the variation of intracavity photon number $\Delta P$ with respect to the photon number at the equilibrium position $P_0$. Within the linear regime \cite{yariv2006photonics} these two quantities are related to the readout voltage by $\Delta P / P_0 = \Delta V / V_0$.
To estimate the sensitivity of our setup we roughly evaluated the numerical solution of Eq. \ref{eqn:detoma} for displacements from 0 to 1 $\mu$m around the initial position. Given a readout voltage $V_0 + \Delta V(t)$, the predicted linear displacement is
\begin{equation}
\Delta x_{theo}(t) \simeq 9.76 \cdot 10^{6}[\text{pm}]\frac{\Delta V(t)}{V_0}.
\label{eqn:theory}
\end{equation}
It is important to stress that the linear operation regime we found is valid only if there is no mode competition inside the cavity. If that is not the case, $T_L$ is not well defined anymore, and the relative intensity noise rises sensibly \cite{tromborg1997mode,detoma2005complex}.

\begin{table}
\caption{\label{tab:param}Parameters employed to solve the theoretical laser equations in section \ref{sec:littrow}. The notation, the value and a brief description of the parameters are reported in the three columns.}
\centering
\begin{tabular}{|l|l|l|}
\hline
Parameter&Value&Description\\ \hline
$I_{th}$ & $33\,\text{mA}$ & threshold current\\
$I$ & $63\,\text{mA}$ & current bias \\
$V_c$ & $2.15\cdot 10^{-10}\,\text{cm}^3$ & volume of the active region\\
$\tau_{in}$ & $16\, \text{ps}$ & active region round-trip time\\
$q$ & $-1.6\cdot 10^{-19}\,\text{C}$ & carrier elementary charge\\
$\tau_s$ & $1.4\,\text{ns}$ & carrier lifetime\\
$N_{th}$ & $\tau_s I_{th}/q V_c$ & threshold carrier density\\
$r_2$ & $0.07$ & diode left facet reflectivity\\
$r_g$ & $0.8$ & grating maximum reflectivity\\
$\Delta \omega$ & $126\cdot 10^9\,\text{rad}\cdot\text{s}^{-1}$ & grating spectral linewidth\\
$\omega_G$ & $2\pi f_L$ & grating central frequency\\
$r_G$ & $rg/(1+j(\omega-\omega_G)/\Delta\omega)$ & grating approximated reflectivity\\
$L$ & $1.0\,\text{cm}$ & grating distance from diode left facet\\
$\tau$ & $2L/c_0$ & grating to left diode facet round-trip time\\
$r_{ext}$ & $0.3$ & Si$_3$N$_4$ membrane reflectivity\\
$L_{ext}$ & $20\,\text{cm}$ & Si$_3$N$_4$ membrane distance from diode left facet\\
$\tau_{ext}$ & $2L_{ext}/c_0$ & Si$_3$N$_4$ membrane to diode left facet round-trip time\\
$c_0$ & $3\cdot 10^8 \, \text{m}\cdot \text{s}^{-1}$ & speed of light in vacuum\\
\hline
\end{tabular}
\end{table}

\subsection*{\label{sec:calibration}Connection between readout voltage and displacement: experimental calibration}
Using eq. \ref{eqn:theory} we can link the measured voltage to the membrane displacement. To verify that our model well reproduces the experimental conditions, we carried out a calibration of the system by employing a commercial atomic force microscope (AFM). 
As the cantilever of the AFM has a mass comparable to the Si$_3$N$_4$ membrane and a resonance frequency of $\sim 80 \,\text{kHz}$, it is not possible to follow the motion of the membrane without strongly perturbing it. Therefore we chose to calibrate the motion using a mirror with the same reflectivity of the membrane, but deposited directly on the chip nitride film. The piezoelectric actuator was driven with a sinusoidal tone at frequency $f_0$ and amplitude ranging from 10 mV$_{\text{rms}}$ to 2 V$_{\text{rms}}$. The bulk displacement was then recorded with the AFM probe. We then measured the voltage readout obtained including the same mirror in the optomechanical setup, with the same driving conditions. The two sets of data are linearly proportional to the piezo drive, therefore we correlated the measured displacements with the voltage readout to get a calibrated setup sensitivity:
\begin{equation}
\Delta x_{cal}(t) = 2.6 \pm 0.3\, [\text{pm}\cdot\text{nV}^{-1}] \Delta V(t).
\label{eqn:calibration}
\end{equation}

\section*{Results and discussion}

\subsection*{\label{sec:p_sweep}Air pressure effect on the mechanical spectrum}
\begin{figure}
\includegraphics[width = \linewidth]{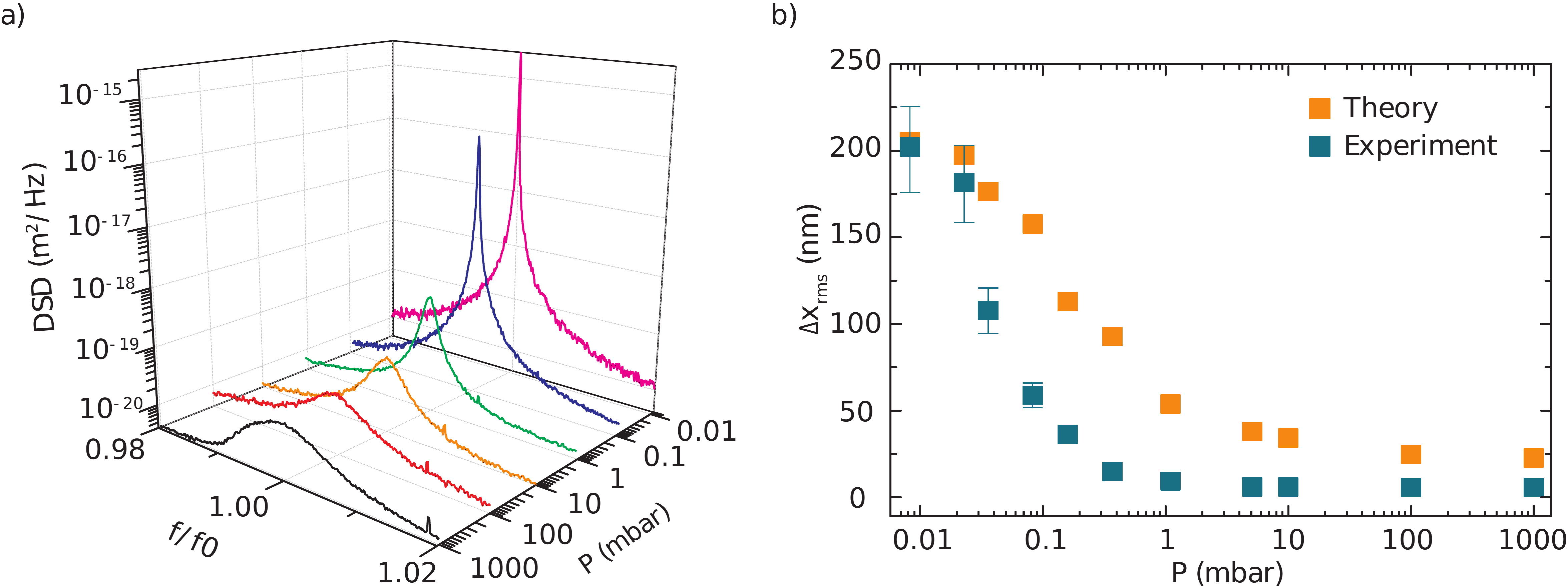}
\caption{\label{fig:p_sweep}a) Displacement spectral density as a function of the air pressure. Note that the lineshape gets slightly distorted close to atmospheric pressure due to mechanical nonlinearities. b) Extracted total displacement compared with the linear theory prediction.}
\end{figure}
We first characterized the effect of the air pressure on the motion of the suspended mirror. The membrane was placed inside a vacuum chamber, where the internal pressure was changed from $1\, \text{bar}$ to $8.3 \cdot 10^{-3}\,\text{mbar}$. The motion of the mirror was forced by a piezoelectric ceramic actuator, driven by a flat spectrum voltage defined as:
\begin{equation}
V_D (f) = 10\,\text{mV}_{\text{rms}}, \quad
\text{through the whole frequency span}.
\label{eqn:drive}
\end{equation}
At different pressures, we measured the spectrum of $\Delta V (t)$ in a homodyne scheme, by employing the piezo driving source as the local oscillator. The measurement resolution bandwidth (RBW) was of $1.56$ Hz, spanning over a 5 kHz spectral range centered on $f_0$.

Using Eq. \eqref{eqn:calibration} the power spectral density is translated into the calibrated displacement spectral density ($DSD$), which measurements are reported in Fig. \ref{fig:p_sweep} (a). The amplitude spectrum at $1 \,\text{bar}$ is clearly asymmetric, denoting a nonlinear dynamics of the oscillator; symmetry is gradually recovered as the pressure decreases.
This behaviour is confirmed if we compare the total rms displacements $\Delta x_{rms}$ extracted from the spectra with the solutions of a classic forced linear harmonic oscillator, as shown in Fig. \ref{fig:p_sweep}(b). The driving source has a fixed displacement spectrum, in order to resemble the effect of the piezo actuator, while $f_0$ and the damping rate are taken from the experimental data. Experiment and theory show a good agreement at low pressures; on the other hand, a systematic mismatch is present at high pressures, with the theoretical estimate higher than the experimental data. This can be attributed to the increased air friction, expressed by the coupling of the fundamental mode of oscillation with the air modes in the box underneath the mirror, and the squeeze film effect of the air box beneath the membrane, which increases the damping rate and the effective mass of the vibrational mode \cite{blech1983isothermal,bao2007squeeze}.
For pressures lower than $2\cdot 10^{-2}\,\text{mbar}$, both the oscillation amplitude and the $Q$ factor tend to saturate, the latter assuming a value of $Q = 3.4 \pm 0.3 \cdot 10^4$, which we believe is still partially limited by thermoelastic damping \cite{reinhardt2015tramp}.

\subsection*{\label{sec:piezo}Thermal and piezo-driven spectra}
\begin{figure}
\includegraphics[width= \linewidth]{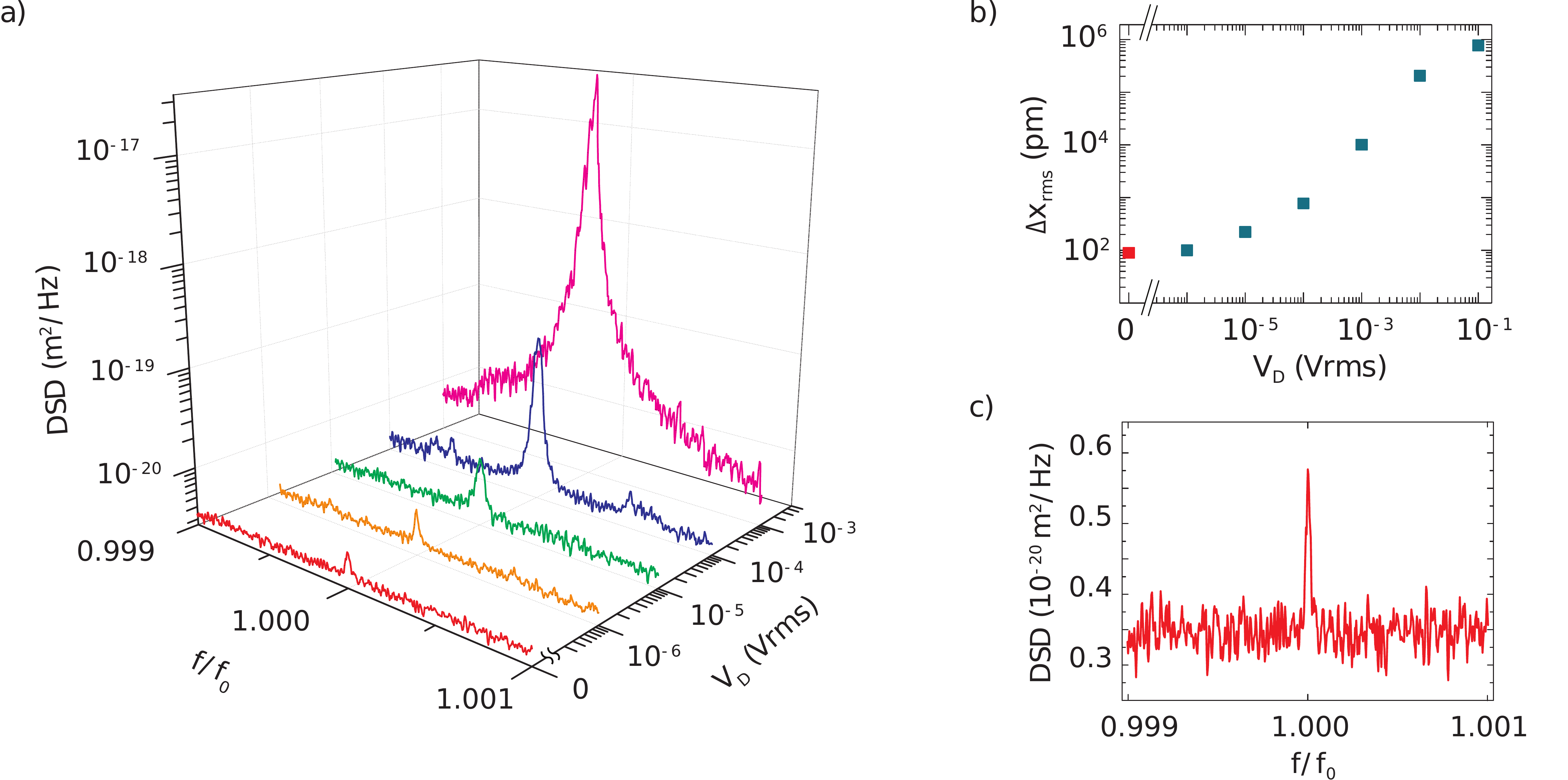}
\caption{\label{fig:piezo}a) Displacement spectral density as a function of the piezo driving voltage $V_D$. b) Extracted total rms displacement of the membrane $\Delta x_{rms}$ as function of the piezo driving voltage $V_D$. b) Displacement spectral density resulting from the thermal fluctuation of the membrane.}
\end{figure}
At the lowest achievable pressure ($8.3 \cdot 10^{-3}\,\text{mbar}$ ), we focused on the displacement sensitivity of the apparatus, measuring the displacement spectra at different driving voltages of the piezo actuator.
To increase the spectral resolution, the signals were acquired with a 250 mHz RBW, spanning over a 200 Hz spectral window centered on $f_0$. The piezo driving voltage had the same functional shape described in eq. \eqref{eqn:drive}, but its amplitude was varied in a range between $1\,\mu\text{V}_{\text{rms}}$ and $100\,\text{mV}_{\text{rms}}$. Figure \ref{fig:piezo} (a) shows the measured $DSD$ for different driving voltages down to $V_D=0$. The $\Delta x_{rms}$ has been reported in panel (b). For large values of $V_D$ it does not vary linearly with the piezo voltage; this can be due to a nonlinearity in the emitted power arising from mode competition inside the cavity, or to the breakdown of the linear elastic regime caused by the strong bending of the tethers which reduces the oscillation amplitude. As $V_D$ is decreased, the linear behavior is recovered, and $\Delta x_{rms}$ converges to the displacement induced by Langevin thermal fluctuations. By switching off $V_D$ and directly acquiring the voltage signal, the thermal peak from the membrane is clearly visible, as reported in Fig. \ref{fig:piezo} (c). Using again the results of our calibration (eq. \eqref{eqn:calibration}), we estimate a thermal displacement of $\Delta x_{\mathbb{T},cal} = 87 \pm 11 \,\text{pm}$. This has to be compared with the theoretical feedback evaluation, which gives a displacement of $\Delta x_{\mathbb{T},theo} = 59.7 \,\text{pm}$, where eq. \eqref{eqn:theory} has been used on the integrated voltage spectrum ($\Delta V = 33.6 \, \text{nV}$, $V_0 = 5.5 \, \text{mV}$). Both experimental and theoretical results are in good agreement. By simply considering the phonon population at such temperature, we can have another rough theoretical estimate of the expected RMS thermal fluctuations by considering the equipartition theorem. The numerical simulations performed with a thermal transport finite-element-method solver (COMSOL Multiphysics) predict that the laser beam heats up the mirror to an average temperature of $T=354\,\text{K}$. The resulting thermal fluctuation is
\begin{equation}
\Delta x_{\mathbb{T},EQ}=\sqrt{\frac{2 k_B T}{4\pi^2 m_{eff} f_0^2}}\simeq 38.3 \, \text{pm},
\label{eqn:kt}
\end{equation}
where $k_B$ is the Boltzmann's constant and $m_{eff}=3.141\cdot 10^{-11}\, \text{Kg}$ the mechanical mode effective mass. Even in this case we have a reasonable agreement with the experimentally obtained displacement. 

In conclusion, we employed a compound cavity laser diode for measuring small displacements of a silicon nitride trampoline membrane down to the thermal noise in the linear regime. The membrane movement directly shows in the laser emission thanks to the effect of optical feedback. By measuring average displacements as low as tens of picometers, our system can represent a fast, compact and cost effective optomechanical platform. Further upgrades of the setup, such as an improved environmental filtering and thermal and mechanical stabilization of the optical components, would push the resolution down to few pms, allowing for measurement of thermal motion at few K temperatures ($\Delta x_{\mathbb{T}}\sim 10\,\text{pm}$ at $30\,\text{K}$). Moreover, the added functionality of getting a feedback directly on the laser source will enable the use of strong optical nonlinearities which can interact with mechanical elements through radiation pressure. In this active cavity optomechanical systems, schemes in which optical chaos is ingrained in a mechanical state can be readily imagined, opening the route to the investigation of a new class of mesoscale non-Markovian phenomena.

\section*{Methods}

\subsection*{Sample fabrication and experimental setup}
The trampoline membranes have been fabricated starting with a 300nm LPCVD stoichiometric silicon nitride film grown on a $250\, \mu \text{m}$ thick silicon FZ 2'' wafer. The high temperature at growth ($\sim 800 ^{\circ}C$) and the different thermal expansion between Si and Si$_3$N$_4$ gives the thin film a considerable tensile stress at room temperature ($\sim 900 \, \text{MPa}$).
At first the metallic mirrors were defined through e-beam lithography followed by a 5/50 nm Ti/Au thermal deposition and lift-off. A second aligned beam-write defined the membrane pattern which was transferred on the sample with plasma etching (CF$_4$/H$_2$). Finally the full membrane was released in a hot KOH solution.
The membrane was mounted on a flat piezoceramic actuator from PI with a fundamental resonant frequency of $300 \, \text{kHz}$. The full block was then placed in a vacuum chamber with optical access and carefully aligned with the laser optical path. Sample alignment was performed by using not-suspended mirrors placed on the same chip which were shaken by the piezo actuator. The drive voltage was a flat band voltage coming from the source port of a Hewlett Packard 89441A vector signal analyzer, which collected the drive voltage at port 1 and the readout voltage by the laser integrated photodiode at port 2. The scattering coefficient $S_{21}$ was then maximized to obtain a good optical alignment of the system.

\subsection*{Theoretical modeling, simulations and calibration}
Equation system \eqref{eqn:detoma} in the main text was numerically solved by using the commercial software Mathematica, with semi-empirical initial conditions and parameters independently extracted from the experiment. The membrane motion was simulated using a commercial FEM solver (Comsol Multiphysics) in order to predict the membrane frequency and motional mass. The simulation prediction was then independently confirmed by using a Polytech UHF-120 laser vibrometer.
The calibration of the setup was performed with a Bruker Dimension Icon AFM. The deflection signal coming from the cantilever, in contact with the sample surface, was recorded in time using a high-speed detector ($50 \, \text{MHz}$ bandwidth) following the Thermal Tune calibration protocol.

\bibliography{bib}

\section*{Acknowledgements}

The authors wish to acknowledge fundings from the ERC Advanced Grant SOULMAN (ERC-FP7-321122) and QUANTOM project of the italian MIUR. D.N.U. acknowledges Ramon y Cajal Contract from spanish MINECO.

\section*{Author contributions statement}
A.T. conceived and supervised the whole work. L.B. and A.P. conducted the main experiment and analysed the results, L.B. performed the theoretical calculations, A.P. fabricated the sample. L.B., A.P., L.M. and D.N.U. devised the setup, L.B., A.A. and F.C. performed the calibration. L.B. and A.P. wrote the manuscript. All authors reviewed the manuscript. 

\section*{Additional information}

\textbf{Competing financial interests} 
The authors declare no competing financial interests.

\end{document}